\documentclass[aps,prl,twocolumn,superscriptaddress,showpacs,
preprintnumbers,amsmath,amssymb,floatfix]{revtex4-1}
\usepackage{epsfig}
\usepackage{graphicx}
\usepackage{color}
\usepackage{amsmath}
\usepackage[english]{babel}
\usepackage{amssymb}
\usepackage[normalem]{ulem}

\begin{document}

\def\bc{\begin{center}} 
\def\ec{\end{center}}
\newcommand{\beq}{\begin{equation}}
\newcommand{\eneq}{\end{equation}}
\newcommand{\beqnn}{\begin{equation*}}
\newcommand{\eneqnn}{\end{equation*}}
\newcommand{\beqy}{\begin{eqnarray}}
\newcommand{\eneqy}{\end{eqnarray}}
\newcommand{\beqynn}{\begin{eqnarray*}}
\newcommand{\eneqynn}{\end{eqnarray*}}
\newcommand{\proj}[1]{\ket{#1}\bra{#1}}
\newcommand{\ket}[1]{\left |{#1}\right \rangle}
\newcommand{\braket}[2]{\langle #1 | #2\rangle}
\newcommand{\bra}[1]{\langle #1 | }

\newcommand{\inprod}[2]{\braket{#1}{#2}}
\newcommand{\Tr}{\mathrm{Tr}}
\newcommand{\vp}{\vec{p}}
\newcommand{\Or}{\mathcal{O}}    
\newcommand{\so}[1]{{\ignore{#1}}}

\newcommand{\red}[1]{\textcolor{red}{#1}}
\newcommand{\blue}[1]{\textcolor{blue}{#1}}
\newcommand{\green}[1]{\textcolor{green}{#1}}

\title{Generalized quantum microcanonical ensemble from random matrix product states}

\author{Silvano Garnerone}		
\affiliation{Institute for Quantum Computing, University of Waterloo, Waterloo, ON N2L 3G1, Canada}

\author{Thiago R. de Oliveira }
\affiliation{Instituto de Fisica, Universidade Federal Fluminense,
Av. Gal. Milton Tavares de Souza s/n, Gragoata 24210-346, Niteroi, RJ, Brazil}

\begin{abstract}
We propose a tensor network algorithm for the efficient sampling of 
quantum pure states belonging to a generalized microcanonical ensemble. 
The algorithm consists in an adaptation of the power method to a recently 
introduced ensemble of random matrix product states. The microcanonical ensemble 
that we consider  
is characterized by the fact that the participating energy eigenstates  
are not required to have identical statistical weight. To test the method we apply it to
the Heisenberg model with an external magnetic field, and we find that 
the magnetization curves, due to the microcanonical constraint, are qualitatively different 
from those obtained in the canonical ensemble. 
Possible future applications include the study of isolated quantum systems evolving after a quantum quench.
\end{abstract}


\maketitle
{\it Introduction}.---
{The problem of representing statistical mechanics ensembles within the more fundamental quantum theory
has recently attracted new interest both at the experimental and theoretical level} 
\cite{kinoshita_quantum_2006,hofferberth_non-equilibrium_2007,Bloch2008,Lewenstein2007,gemmer2004quantum,polkovnikov_colloquium:_2011,
Popescu2006,Goldstein2006,reimann_typicality_2007}. 
On the experimental side, the ability to isolate quantum systems from 
their surrounding triggers new questions 
on quantum thermalization and quantum equilibration dynamics \cite{rigol_thermalization_2008}. 
On the theoretical side, a conceptually new approach based on quantum typicality   
has emerged as an alternative paradigm 
for the understanding of quantum statistical mechanics and thermodynamics 
\cite{Popescu2006,Goldstein2006}.
Since in quantum mechanics ensembles are represented by density matrices, there are two 
ways in which they can be obtained: 
(i) after a partial trace over an environment {which is entangled with the system in a global pure state}, 
(ii) averaging over an ensemble of {global} pure states. 
The paradigmatic change provided by quantum typicality {has taken advantage} 
of the {partial trace point of view}. 
{Loosely speaking quantum typicality} 
exploits the concentration, around {the} average {value},  
of some random variable in high dimensional spaces. {This} can be used 
{for example} to justify 
the occurrence of the canonical ensemble 
as the reduce{d} density matrix 
of a typical random pure state 
\cite{Popescu2006,Goldstein2006a,reimann_typicality_2007,Reimann2008}.
On the other hand, averaging over ensembles of microscopic 
configurations --point (ii) above--is yet another effective 
way of representing properties of physical systems \citep{Goldstein2006,white_minimally_2009}. 
In {this} work we adopt this second approach, 
and we provide a tensor network algorithm 
{to} sample 
from a generalized quantum microcanonical ensemble, 
which differs from the standard microcanonical construction because the requirement 
for the participating eigenstates to have  
equal statistical weights is relaxed to a more realistic one \cite{reimann_typicality_2007}. 
The algorithm consists in an adaptation of the power method {to Matrix Product States (MPS)}, 
and it exploits statistical properties of 
a recently introduced ensemble of Random MPS (RMPS) 
\cite{Garnerone2010,Garnerone2010a}, which is the key feature allowing for the 
computational efficiency of the method. In fact, any alternative procedure which 
considers Haar-distributed random pure states  
would be far less efficient than the present scheme, due to the fact that 
those states are computationally hard to generate. 
For the Heisenberg model with an external magnetic field, 
the qualitative comparison of the results of our simulations 
with a different approach exploiting Haar-distributed states, 
also supported by exact calculations 
\cite{sugiura_thermal_2012}, corroborates the validity of our method. 
The introduced computational procedure, 
and the relative generalized microcanonical ensemble, can also be of relevance in the context 
of current experimental setups where the quantum system is almost isolated from the environment 
\cite{kinoshita_quantum_2006,hofferberth_non-equilibrium_2007,Bloch2008}.    

{\it The RMPS ensemble}.---
MPS are quantum pure states fully specified 
by a set $\{ A^{\sigma_i}, i=1,\dots,N \}$ of relatively small dimensional matrices \cite{Schollwock2011}:
\beqy 
\ket{\psi}&=&\sum_{\boldsymbol{\sigma},\boldsymbol{i}} A_{i_1,i_2}^{\sigma_1}A_{i_2,i_3}^{\sigma_2}\cdots 
A_{i_{N-2},i_{N-1}}^{\sigma_{N-1}}A_{i_{N-1},i_N}^{\sigma_{N}} \nonumber\\
&&\times \ket{\sigma_1 \sigma_2 \cdots \sigma_{N-1}\sigma_N}.
\label{eq:rmpsket} 
\eneqy
Following the standard notation for open-boundary MPS  
we set the row and column indices $\{i_1,i_N\}$ to $1$; while all other indices 
belong to the set of integers from $1$ to $\chi$, the latter being the so called bond-dimension of the MPS. 
For two levels systems, it follows that 
an MPS is specified by no more than $2N \chi^2$ numbers 
which, for $\chi$ not too large, is exponentially smaller than the $2^N$ 
parameters required by a typical quantum state in the same Hilbert space $\mathcal{H}$.
Indeed, any state can be written as a MPS for $\chi$ {large enough}, 
but the advantage in using MPSs occurs only when $\chi \sim {\rm poly}(N)$. 
Moreover, MPS are physically relevant since they are good approximations of ground states
of short-ranged gapped Hamiltonians, 
and they can also be used to construct thermal states \citep{Schollwock2011,white_minimally_2009,Stoudenmire2010}. 
With the aim to better understand the usefulness of typicality in
quantum statistical mechanics,  in \cite{Garnerone2010} 
an ensemble of random MPS has been proposed    
{sharing} some of the statistical
features of Haar-distributed random pure states.
According to the construction presented in \cite{Garnerone2010}, 
a random MPS is defined considering 
the $\chi$-dimensional matrices $A^{\sigma_i \in \{\uparrow,\downarrow\}}$ 
as contiguous sub-blocks of a set of $N$ independent Haar-distributed unitaries $U_i$,  
each unitary having dimension $2\chi$ (where $2$, in general, 
is the dimension $d$ of the local Hilbert space).
In particular, a sufficient condition for the occurrence of typicality in the case of RMPS has been derived in \cite{Garnerone2010}, 
and the exact calculation of the average density matrix of the ensemble has been 
obtained in \cite{Garnerone2010a}. Interestingly enough, the 
average random MPS coincides with the one 
obtained from Haar-distributed random states, i.e. it is the maximally mixed state.
For an alternative construction of random physical states, 
though not related to MPS, see also \cite{Hamma2012,Hamma2012a,Zanardi2013}.
\begin{figure}
\includegraphics[scale=0.35]{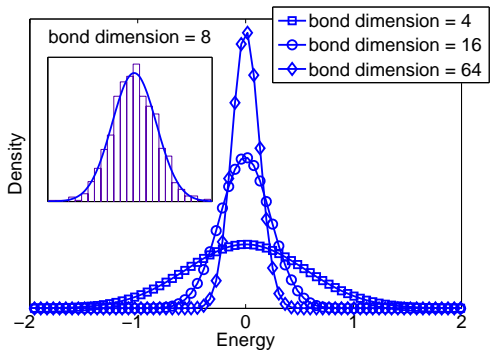} 
\caption{Gaussian fits for the histograms of the expectation values of the 
transverse field Ising chain with respect to RMPSs. For ease of exposition, in the main 
figure only the normal fitting curves are shown, while 
the inset shows a particular histogram with the relative Gaussian fit, for $\chi=8$.
$N=10, \chi=\{4,16,64\}, 1000$ realizations. }
\label{fig:figure1}
\end{figure}
In this letter we show how the RMPS ensemble, together with efficient 
and relatively simple computational techniques,  
can be used to sample from a generalized quantum microcanonical ensemble introduced
by Reimann \cite{reimann_typicality_2007}. 
Computational efficiency is inherited from well-known 
MPS features, while on the statistical side {we also}  
exploit some of the properties characterizing the RMPS ensemble. In particular   
we will use the fact that the average state is equal to the 
completely mixed state $\left[ \proj{\psi} \right]_{\rm{ave}}={\mathbb{I}}/{D}$, 
where $\ket{\psi}$ is a random MPS as in Eq.\ref{eq:rmpsket}, and $D$ is the dimension of $\mathcal{H}$. 
One of the implications of the above identity is numerically checked  in Fig.\ref{fig:figure1} , where 
the expectation value {of the energy} of the transverse field Ising chain
is sampled over 1000 realizations of independent random MPS.
The  results of the simulations confirm that, irrespective of the auxiliary bond 
dimension $\chi$, the average energy is given by 
$\left[ \bra{\psi}H\ket{\psi}\right]_{\rm{ave}}={\Tr{H}}/{D}=0.$
Moreover in Fig.\ref{fig:figure1} one observes  that the variance 
of the distribution depends on $\chi$. 
We {are not able} to derive an analytic expression for the second moment of the 
distribution of RMPS, which would provide the variance in the plots of Fig.\ref{fig:figure1}. 
{Nevertheless}, we can check numerically that the variance of the energy 
distribution decreases as a power-law with increasing $\chi$, consistently with typicality
properties of the ensemble of RMPS \citep{Garnerone2010}. 
In the case of Haar-distributed random states 
the characterization of the probability distribution for general observables 
can be found in \cite{Venuti2012,Gallay2011,Puchala2012,Dunkl2011,Dunkl2011a,Brody2007,brody2007quantum}.
{We shall now present} a computational procedure 
allowing us to modify the energy distribution from 
the initial Gaussian centred in $0$ to another normal distribution 
with a desired average value, and whose 
variance can be reduced with a good degree of control. 

{\it A generalized microcanonical ensemble from RMPS}.---
Our goal is to modify the initial distribution of Hamiltonian expectation values 
in such a way that we can arbitrarily 
choose its center $E$, and further decrease its variance $\gamma$. 
To accomplish this we use an iterative technique which, starting 
from an initial RMPS $\ket{\psi}_0$, prepares a final MPS $\ket{\psi}_{\rm{f}}$ 
whose population $p_i\equiv|\bra{E_i}\psi \rangle_{\rm{f}}|^2$ -- with respect 
to the eigenstates $\ket{E_i}$ of the Hamiltonian -- 
is concentrated on the eigenvectors close in energy to a given value $E$. 
The iterative technique is provided by {the} power method: 
consider an operator $G$ with maximum eigenvalue $\lambda$ 
such that $G\ket{\lambda}=\lambda\ket{\lambda}$. 
{It} is easy to show that any initial state $\ket{\psi}_0$, which is not orthogonal to $\ket{\lambda}$, can 
be used as a starting point to obtain a good approximation to $\ket{\lambda}$, simply iterating the following operation 
\beq
\ket{\psi}_{\rm{k+1}}=\frac{G\ket{\psi}_{\rm{k}}}{\Vert G\ket{\psi}_{\rm{k}} \Vert_2}, \; \{k=0,1,..,{\rm f}{-}1\}. 
\label{eq:iterate}
\eneq 
For our purposes we  set 
$
G=\mathbb{I}-\left(\frac{H - E}{\sigma}\right)^2,
$
where $E$ is a specified energy value, and $\sigma$ is a parameter 
which depends on the spectral range of the {local} Hamiltonian $H$, and which has to be chosen so that $G$ 
is positive semi-definite (a simple estimate is provided 
by $\sigma=2N\delta + \vert E \vert$, 
where $\delta$ is the greatest absolute value of the Hamiltonian parameters).  
The power method will bring an initial RMPS $\ket{\psi}_0$ 
closer to the eigenstate $\ket{E}$, satisfying $H\ket{E}=E\ket{E}$. 
{Indeed}, after a number $k$ of iterations $\ket{\psi}_0$  becomes
$
\ket{\psi}^{(k)}={G^k \ket{\psi}_0}/{\Vert G^k \ket{\psi}_0 \Vert_2}.
$
In the following we provide a simple argument showing that 
the asymptotic distribution of energies is approximated by a Gaussian centred in $E$.
From the last equation we can write the average density matrix at the $k$-th iteration as follows: 
$
\left[\proj{\psi}\right]_{\mathrm{ave}}^{(k)} \propto \left[G^k \proj{\psi} G^k \right]_{\mathrm{ave}} 
= G^k \left[ \proj{\psi} \right]_\mathrm{ave} G^k \propto G^{2k},
$
{where we assumed that with high probability 
the denominator does not significantly depend on the initial state 
(as it occurs for Haar-distributed states \cite{reimann_typicality_2007}), and 
we used the explicit expression for the average state}. 
In the limit of many iterations we can write: 
$G^{2k} \overset{k\gg 1}\sim  \sum_{\vert\frac{E_i-E}{\sigma}\vert \ll 1} \left[ 1-\left(\frac{E_i-E}{\sigma}\right)^2  \right]^{2k} \proj{E_i} 
\sim  \sum_{\vert\frac{E_i-E}{\sigma}\vert \ll 1} \exp{\left[-2k\left(\frac{E_i-E}{\sigma}\right)^2 \right]}\proj{E_i}$,
where the sum is restricted to those eigenvalues sufficiently close to $E$. 
From the above argument one can derive a polynomial upper-bound (in the size of the system) 
on the number $k$ of iterations needed in order to have a sufficiently small variance 
(given by $4\sigma^2/k$): $k \sim \sigma^2 \propto N^2$. 
For any practical purposes though, our numerical simulations show that already one hundred
iterations are sufficient to obtain energy distributions that are concentrated enough, 
and that can be used to analyse the system.
\begin{figure}[!h]
\includegraphics[scale=0.4]{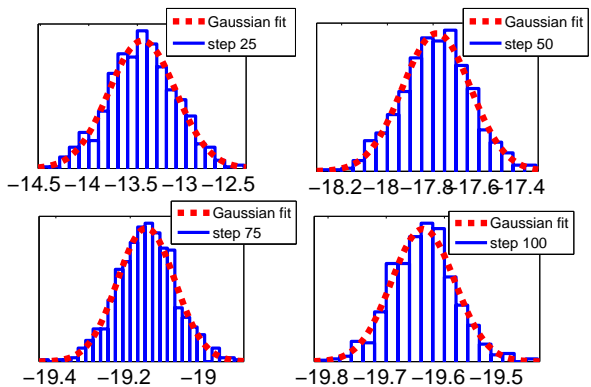} 
\caption{(Color online). Energy histograms and their normal fitting curves 
at different steps of the power-method iterations. 
$N=50, \; \chi=32, \; E=-20, \; 1000$ realizations.}
\label{fig:figure2}
\end{figure}
Fig. \ref{fig:figure2} shows the histograms of energy values obtained at different steps 
of the iteration. The 
simulation confirms that the distribution is indeed a Gaussian 
whose mean approaches asymptotically the value $E$, and whose variance decreases 
algebraically with the number of iterations. 
{The standard Dirac delta function representation of the microcanonical ensemble 
is approached in the limit of many iterations}:   
$\left[\proj{\psi}\right]_{\mathrm{ave}}^{(k)} \overset{k\rightarrow \infty} \longrightarrow \delta(H-E)$. 
Also, note that Ref. \citep{long_finite-temperature_2003} 
provides a similar but more rigorous derivation of the average state in the context of 
Haar-distributed random states,  showing that indeed the asymptotic final distribution 
of energies is a Gaussian centred in $E$. 
For small systems, using exact diagonalization we observe that at the level of a single RMPS realization the populations $p_i$ of energy eigenstates, which 
at the beginning is spread over a wide range of energy values, gets more and more concentrated around the chosen 
$E$ value during the iteration process. 
This is consistent with the characterization of a generalized microcanonical ensemble,
as proposed by Reimann in \citep{reimann_typicality_2007}.
In particular, the standard requirement for the microcanonical ensemble, {i.e.} 
that {\it any} {participating} eigenstate {has the {\it same} probability in} the ensemble, 
is weakened to more general {\it non-uniform} distributions of eigenstates in the {small} energy window, 
like those that we obtain. Reimann argues that these generalized microcanonical 
energy distributions have to be considered more realistic realizations of the quantum microcanonical ensemble 
\cite{reimann_typicality_2007}. 
Indeed, for a not too small system, energy levels are extremely close to each other
 and experimentally it would be extremely hard to prepare an 
 equal superposition of energy eigenstates. 
 On the other hand it {should be} much simpler to require a 
 distribution of the energy population with a sharp peak and a small variance 
(see also \citep{Goldstein2006a} and references therein).
Note that, in principle, for sufficiently {many iterations}, one single realization of a random MPS 
can provide an instance of the quantum microcanonical ensemble, along the lines of the 
construction {presented} in \citep{sugiura_thermal_2012} for Haar-distributed states.
On the other hand, we can exploit the fact that sampling over many RMPS improves 
the estimate of averaged quantities \cite{Hoeffding1963}, 
and we do not need to wait for many iteration steps to {terminate} as long as 
we allow for a small finite width in the distribution of energies around 
their mean value. 
From this point of view the kind of generalized microcanonical ensemble that 
we obtain is provided by an ensemble of random pure states which, 
when expressed in the energy eigenbasis of the Hamiltonian, have support on an energy 
window whose width can be controlled 
by the number of iterations, while the number of sampled states controls the 
accuracy in the estimation of averaged quantities.
Moreover, the distribution obtained by sampling the Hamiltonian expectation values 
is not only meaningful from a statistical mechanics point of view in the regime of {many} iterations. 
In fact, the fluctuations in the energy value obtained at finite and not too large 
$k$ can be interpreted as an energy exchange between the sampled system and a small
 finite-size environment, weekly interacting with the system. 
More about this interpretation can be found in \citep{Challa1988} 
where the so-called Gaussian ensemble has been analysed in the context of classical statistical mechanics. 
For an alternative construction of a generalized quantum microcanonical ensemble see also \citep{Muller2011,Fine2010},  
though in these works there is no constraint on the fact that, in the eigenenergies basis, 
the states in the ensemble have support only in a 
small energy window, and in general they do not reproduce the standard canonical ensemble for subsystems.

\begin{figure}
\includegraphics[scale=0.28]{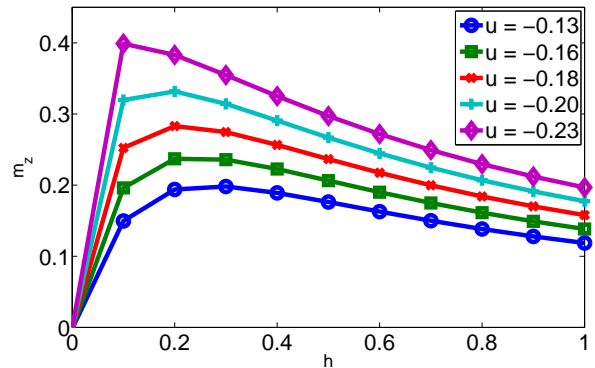} 
\caption{(Color online). Magnetization curves of the 
ferromagnetic Heisenberg chain with an external field, for different values of the energy density. 
$N=50,\chi=16$.
}
\label{fig:figure3}
\end{figure}

\begin{figure}
\includegraphics[scale=0.4]{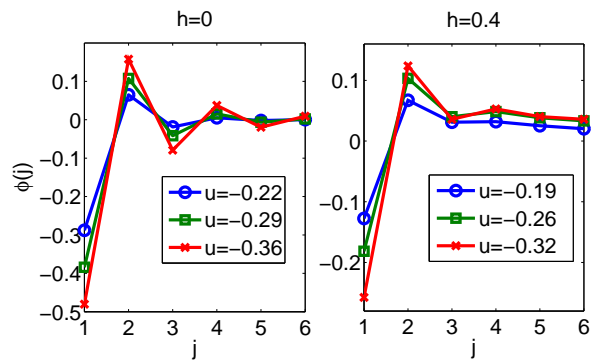} 
\caption{(Color online). Average correlation function of the antiferromagnetic Heisenberg chain 
with an external field, for different values of the energy density. 
The figure on the left shows the results with $h=0$, while 
the figure on the right shows the results for $h=0.4$. $N=50,\chi=16$.}
\label{fig:figure4}
\end{figure}

{\it Application to spin chain Hamiltonians}.---
In this section we apply the RMPS power method to the 
Heisenberg model with an external magnetic field:
$
H = - \sum_{i=1}^N \frac{J}{4}\left(\sigma^x_i \sigma^x_{i+1} + \sigma^y_i \sigma^y_{i+1} + \sigma^z_i \sigma^z_{i+1} \right) +h \sigma^z_i. 
$
To check the validity of the method we compare our results with similar quantities 
calculated {for smaller systems} in \cite{sugiura_thermal_2012}.
We compute the magnetization $m_z\equiv N^{-1}\sum_{i=1}^N \left[ \langle\sigma^z_i\rangle
\right]_{\rm{ave}}$ 
for the ferromagnetic model ($J=1$), and the correlation function 
$\phi(j)\equiv N^{-1}\sum_{i=1}^N \left[\langle\sigma^z_i \sigma^z_{i+j} \rangle\right]_{\rm{ave}}$ 
for the antiferromagnetic model ($J=-1$). 
Both Fig. \ref{fig:figure3} and Fig. \ref{fig:figure4} 
show lines labelled by different energy densities $u\equiv E/N$, and  
each point on the curves is obtained averaging over $200$ 
statistically independent RMPS (we checked that the standard deviations 
are always very small, of the order of $10^{-3}J$). 
We note that, from a computational point of view, 
in order to sample over many random MPS 
the algorithm can be trivially  parallelized to efficiently estimate statistical properties 
of the ensemble. 
The results of our simulations are qualitatively and quantitatively very {similar} to those 
provided in \citep{sugiura_thermal_2012}, where 
the authors can also use data obtained from exact 
calculations {in the canonical ensemble} which are consistent with their findings.  
We note that, although our algorithm shares some aspects with the procedure 
presented in \cite{sugiura_thermal_2012}, the two computational schemes 
are substantially different, mainly due to the fact that
we make use of {an MPS ensemble,}  
allowing us to access much larger spin chains.
From a physical point of view it is interesting to note that 
Fig. \ref{fig:figure3} shows a non-monotonic behavior of the magnetization with 
respect to the field, which is quite different from what
one obtains in the canonical ensemble at fixed temperature 
(i.e. monotonic magnetization curves) \cite{Bonner1964}. 
{Note that this difference between the two ensembles 
is evident when looking at the 
`global' behavior of the magnetization curves, while `locally' point-by-point 
one can always find canonical ensembles (characterized by different temperatures) 
providing the same results as the generalized microcanonical one \cite{sugiura_thermal_2012}.}
The difference is due to the energy constraint imposed 
by the quantum microcanonical ensemble, 
which does not allow the system to explore the same set 
of states as in the canonical case. 
In fact there are two main contributions to the Hamiltonian expectation value: 
one due to the interaction term, 
the other due to the magnetic field term. 
For small increasing values of the magnetic field the system 
can increase its magnetization, while keeping the energy constant, 
due to the effect of the interaction term which can balance the different energy 
contribution coming from the magnetic field term 
(this explains the left side in Fig. \ref{fig:figure3}). 
On the other hand, when the field is big enough, a further increase 
in the magnetization can not be compensated by the interaction term, 
implying that the microcanonical energy constraint would not be satisfied. 
The only states allowed in this regime are those for which the magnetization 
decreases with an increasing field. 
Considering state-of-art experiments with almost isolated quantum systems, 
it would be interesting to have an experimental confirmation of this behavior 
{in those setup where it is possible to access the energy window of interest, 
to control the external field 
and to measure the magnetization.}   
Fig. \ref{fig:figure4} provides information on the decay of the correlations and on the way in 
which they can be affected by an external magnetic field. Also in this case our simulations compare well with 
those provided in Ref. \cite{sugiura_thermal_2012}.

{\it Conclusions and future directions}.---
We have presented a tensor network algorithm which can be used to sample 
states belonging to a given energy window. 
Statistical properties of the 
initial ensemble of random MPS, and the exploitation of 
tensor network structures make the algorithm suitable for the simulation 
of large one-dimensional systems.   
This algorithm allows us to test new ideas in {the foundations of} quantum statistical 
mechanics regarding a generalization of the quantum microcanonical ensemble, as 
presented in \citep{reimann_typicality_2007}.
With respect to this, in the future we plan to study in more details computational 
aspects connected to quantum typicality in our ensemble of random MPS.
Our approach suggests also a new algorithm for the simulation of quantum systems in the canonical ensemble, 
with some features similar to the algorithm proposed in \citep{white_minimally_2009}. 
In the context of quantum computation it would also be interesting 
to investigate the 
computational complexity of a quantum algorithm able 
to sample from the microcanonical distribution. 
Along the lines of our construction, one should combine a subroutine implementing a 
quantum pseudo-random circuit \cite{Harrow2009} used to generate random quantum states, 
together with the iterative application of the operator $G$ in Eq.\ref{eq:iterate} using the 
technique developed in \cite{Harrow2009a}. 
Alternatively one could think of adapting a similar quantum circuit 
into the framework of mixed states quantum computation \cite{Knill1998,Ambainis2000}.

{\it Acknowledgements.} SG is grateful to L. Campos Venuti, I. Di Marco, 
S. Haas and P. Zanardi for insightful comments. TRO thanks R. G. Pereira and J. A. H. Neto for useful discussions.
\bibliographystyle{apsrev4-1}
\bibliography{biblio}

\appendix
\section{Generation of random MPS}
{The Hilbert space of a system of $N$ qubits is exponentially big in $N$, 
hence typically it would require an exponential number of coefficients to characterize a state in a given basis.
Matrix product states provide a solution to this problem, since they allow for a compact 
description of the states \cite{Perez-Garcia2007,Schollwock2011}. MPS are represented as linear combinations of basis vectors} whose coefficients  
are given in terms of products of matrices. For a system composed of $N$ particles,
with local Hilbert space dimension $d$, {an MPS} state {can be} specified
{using only a number} $dN$ {of} $A^{\sigma_{i}}$ matrices {(where $i=1,\dots,N$), of dimension at most $\chi$  
(a parameter known as the bond dimension)}. Thus the
$d^{N}$ coefficients {required for a typical state in the Hilbert space} are {compressed into a set of at most}
$dN\chi^{2}$ numbers,
which {makes computations much more efficient, but of course also restricts  the
class of possible states. Fortunately this restricted set can represent good approximations for
many interesting states of relevance in condensed matter. More specifically, MPS are
able to represent any state whose entanglement obeys the area law or only violates it logarithmic.} 

With respect to the boundary conditions, there are two different constructions of MPS: with periodic or with open boundaries.
In the first case the state is written as
\[
|\psi\rangle=\sum_{\boldsymbol{\sigma}}\text{Tr}[A^{\sigma_{1}}A^{\sigma_{2}}...A^{\sigma_{N}}]|\sigma_{1}\sigma_{2}...\sigma_{N}\rangle,
\]
while for open boundaries the first and last matrices are {row}
and {column} vectors

\[
|\psi\rangle=\sum_{\boldsymbol{\sigma}}\langle\phi_{L}^{\sigma_1}|A^{\sigma_{2}}...A^{\sigma_{N-1}}|\phi_{R}^{\sigma_N}\rangle|\sigma_{1}\sigma_{2}...\sigma_{N}\rangle.
\]
One can also consider the case where the $A$ matrices are the same
at each site, {a so-called} homogeneous MPS. Together
with periodic boundary conditions this {also} gives a translational invariant
state. Note that a general non-homogeneous MPS, with or without open boundaries,
can also describe a translational invariant state {\cite{Perez-Garcia2007}}.
There is no unique correspondence between
the set of pure MPS states and the set of $A$ matrices: different
set of matrices may generate the same MPS. This gauge {degree of}
freedom can be fixed using {two} canonical form{s} for
the MPS{: left-canonical MPS or right-canonical MPS}. 
In the {left-}canonical form the $A$ matrices satisfy {the following condition}: $\sum_{\sigma_l}A^{\sigma_l\dagger}A^{\sigma_l}=\mathbb{I}$.
{In the right-canonical form the $A$ matrices satisfy the following condition: $\sum_{\sigma_l}A^{\sigma_l}A^{\sigma_l\dagger}=\mathbb{I}$. 
When constructing an MPS in canonical form the boundary sites can violate the above conditions, but this simply means that 
the state is not normalized and it also provides an efficient way to evaluate its norm \cite{Schollwock2011}}.

{In defining the ensemble of random MPS \cite{Garnerone2010} we took inspiration from the sequential generation of MPS \cite{Schon2007,Perez-Garcia2007}, 
according to which a matrix product state can be seen as the outcome of a sequential interaction of unitary matrices 
between local physical degrees of freedom and an ancillary space. Along this line a natural way for 
defining a random MPS starts by considering a random Haar-distributed unitary matrix U \cite{Mezzadri2007} of dimension $d\chi$, which 
can be seen as composed of blocks in the following way (for simplicity we now restrict to the case $d=2$)
\[
U=\left[\begin{array}{cc}
M & V \\
N & W
\end{array}\right],
\]
where each block has dimension $\chi$. Identifying the block $M$ and $N$ with the two $A^{\sigma_l=1,0}$ matrices associated 
to a local qubit immediately provides the building block for the construction of a left-canonical random MPS. On the 
other hand a right-canonical MPS can be obtained by identifying the two $A^{\sigma_l=1,0}$ matrices with $M$ and $V$. Similar considerations 
hold for the block column matrices $V$ and $W$, or for the row column matrices $N$ and $W$. 
A non-homogeneous random MPS can be iteratively constructed by sampling from a set of $N$ independent random unitary matrices, each 
of dimension $2\chi$.}
In the case of open boundary conditions the first and last {$A$} matrices
are simply {obtained from random Haar-distributed vectors}.
Therefore the {ensemble of random} MPS generated {in this way}  
{is appealing for different reasons: it has a natural operational meaning, 
the state is obtained in a canonical form, and 
it allows to use properties of the Haar distribution. 
Typicality and other properties of the ensemble of random MPS have been studied in \cite{Garnerone2010,Garnerone2010a,Collins2012}}.

\section{Computational cost of the algorithm}
Our algorithm {is an instance of the simple and widely used} power method, an iterative technique {which}
does not {require any diagonalization or singular value decomposition}. 
Instead one has to deal only with matrix-vector multiplications, during which an operator $G$ is 
repeatedly applied to an initial state $\ket{\psi}$. 

In order to estimate the computational resources required by our scheme, 
we start by defining the operations needed in its implementation: generation of 
a random matrix product state $\ket{\psi}$; construction of the operator $G$; application 
of the operator $G$ to $\ket{\psi}$; if needed the evaluation of some expectation values. 
Except for the generation of random MPS, all other computational steps are already well discussed in the 
literature (see for example \cite{Schollwock2011}), and are summarized here just for completeness. 

The generation of the initial random MPS has been described in the previous section. To estimate 
the computational resources, in the worst case, we only need to multiply by $N$ (the length of 
the chain) the cost of the generation of a random unitary of size $d\chi$, where $d$ is the local 
Hilbert space dimension ($2$ in the case of qubits) and $\chi$ is the maximum allowed bond-dimension of the MPS. 
Since $\chi$ is at most a polynomial in $N$, the state can be generated with an amount of 
resources scaling polynomially in $N$.  

The construction of the operator $G\equiv \mathbb{I}-\left(\frac{H - E}{\sigma}\right)^2$--where $H$ is the Hamiltonian 
of the system, while $E$ and $\sigma$ are two given numbers-- involves the Matrix Product Operator (MPO) representation 
of the identity matrix and of the first and second powers of the Hamiltonian. The MPO representation is the 
most effective way of dealing with operators in the MPS framework: it simply consists in the generalization 
to operators of the matrix product decomposition for states (see for example \cite{Schollwock2011} for more details)
\beq
O = \sum_{\boldsymbol{\sigma},\boldsymbol{\sigma'}} W^{\sigma_1,\sigma'_1}W^{\sigma_2,\sigma'_2} \dots W^{\sigma_N,\sigma'_N}\ket{\boldsymbol{\sigma}}\bra{\boldsymbol{\sigma'}},
\eneq
where the bond-dimension of the $W$ matrices is typically small (equal to $5$ or $9$ as we will see). 
The MPO representation of the identity matrix is very simple, while less trivial is the representation of the Hamiltonian. 
An efficient way of constructing it is explained in \cite{Schollwock2011} (see page 142). 
It requires an MPO of bond-dimension $5$ 
for the representation of $H$, while the cost of the representation of $H^2$ can be optimized 
using a bond-dimension equal to $9$.
We note that clearly the construction of the operator $G$ has to be done only once at the beginning of the algorithm. 
Hence the MPO representation of $G$ requires the storage of a number $4N$ of matrices of dimension at most $9$ 
($25$ if one does not optimize the construction of $H^2$). 

The third step in the algorithm consists in the application of the MPO $G$ to the MPS $\ket{\psi}$ which, for 
an MPS of bond-dimension $\chi$ and an MPO of bond-dimension $\chi_W$, costs $\Or{(Nd^2\chi^2\chi_W^2)}$.
Since $G$ is composed by a linear combination of different MPOs, and since after the application of $G$ to $\ket{\psi}$ 
the bond-dimension of the state increases by a factor given by the MPO bond-dimension, we need to compress the resulting 
state to one with a smaller bond-dimension. This can be done using a variational technique costing at most $\Or{(N\chi^3\chi_W^3)}$ \cite{Schollwock2011}.
 
The above operations can be repeated for the desired number of times during the execution of the power method algorithm, and 
one can keep track of the expectation values of relevant observables at a cost which is linear in $N$, at most cubic in $\chi$ and at most quadratic in $\chi_W$ \cite{Schollwock2011}. 

To conclude we note that in order to sample many different initial random MPS the algorithm can be trivially parallelized  
without penalizing the efficiency of the scheme.

\end{document}